\documentclass[runningheads]{llncs}
\usepackage{booktabs}
\usepackage{graphicx}

\begin{document}
\title{COVID-Net US: A Tailored, Highly Efficient, Self-Attention Deep Convolutional Neural Network Design for Detection of COVID-19 Patient Cases from Point-of-care Ultrasound Imaging}
\titlerunning{COVID-Net US}
\author{Alexander MacLean\inst{1}* \and
Saad Abbasi\inst{1} \and
Ashkan Ebadi\inst{2} \and
Andy Zhao \inst{1} \and
Maya Pavlova \inst{1} \and
Hayden Gunraj \inst{1} \and
Pengcheng Xi \inst{3} \and
Sonny Kohli \inst{5} \and
Alexander Wong \inst{1,4}}

%
\authorrunning{A. MacLean et al.}
\institute{University of Waterloo, Department of Systems Design Engineering, Waterloo, ON N2L 3G1,
Canada \\ {*} Corresponding Author \email{alex.maclean@uwaterloo.ca}\and
National Research Council Canada, Montreal, QC H3T 1J4, Canada \and
National Research Council Canada, Ottawa, ON K1K 2E1, Canada \and
Waterloo Artificial Intelligence Institute, Waterloo, ON N2L 3G1, Canada \and
Oakville Trafalgar Memorial Hospital, McMaster University, ON, Canada }

\maketitle   
\begin{abstract}
The Coronavirus Disease 2019 (COVID-19) pandemic has impacted many aspects of life globally, and a critical factor in mitigating its effects is screening individuals for infections, thereby allowing for both proper treatment for those individuals as well as action to be taken to prevent further spread of the virus. Point-of-care ultrasound (POCUS) imaging has been proposed as a screening tool as it is a much cheaper and easier to apply imaging modality than others that are traditionally used for pulmonary examinations, namely chest x-ray and computed tomography. Given the scarcity of expert radiologists for interpreting POCUS examinations in many highly affected regions around the world, low-cost deep learning-driven clinical decision support solutions can have a large impact during the on-going pandemic. Motivated by this, we introduce COVID-Net US, a highly efficient, self-attention deep convolutional neural network design tailored for COVID-19 screening from lung POCUS images. Experimental results show that the proposed COVID-Net US can achieve an AUC of over 0.98 while achieving 353$\times$ lower architectural complexity, 62$\times$ lower computational complexity, and 14.3$\times$ faster inference times on a Raspberry Pi. Clinical validation was also conducted, where select cases were reviewed and reported on by a practicing clinician (20 years of clinical practice) specializing in intensive care (ICU) and 15 years of expertise in POCUS interpretation. To advocate affordable healthcare and artificial intelligence for resource-constrained environments, we have made COVID-Net US open source and publicly available\footnote{https://github.com/maclean-alexander/COVID-Net-US/} as part of the COVID-Net open source initiative.

\end{abstract}
  \vspace{-0.07in}
\section{Introduction}
  \vspace{-0.07in}
Since the beginning of 2020, the Coronavirus Disease 2019 (COVID-19) pandemic has had an enormous impact on global healthcare systems, and there has not been a region or domain that has not felt its impact in one way or another. While vaccination efforts have certainly been shown as effective in mitigating further spread of COVID-19, screening of individuals to test for the disease is still necessary to ensure the safety of public health, and the gold-standard of this screening is reverse transcriptase-polymerase chain reaction (RT-PCR) \cite{ref_pcr}. With RT-PCR being laborious and time-consuming, much work has gone into exploring other possible screening tools, namely using chest x-ray (CXR) and computed tomography (CT) to observe abnormalities in images, and extending from needing to have radiologists analyze the images and make diagnoses to using automated models to screen for the presence of COVID-19 \cite{ref_covid_net,ref_covid_ct}.

Another imaging modality that is gaining traction as a tool for treating lung related diseases is the lung point-of-care ultrasound (POCUS), and has been suggested as most useful in contexts that are resource limited, such as emergency settings or low-resource countries \cite{ref_lung_us,ref_pocus1}. POCUS devices allow for easier and quicker applications than CXR and CT systems, and are much cheaper to acquire leading to more access in a variety of point-of-care locations, thus enhancing the ability for possible COVID-19 screening \cite{ref_pocus1}.

Despite the comparably superior ease-of-use and -access for POCUS devices, the interpretation of ultrasound images is far from simple, and protocols have been written providing instruction for radiologists to standardize such interpretation \cite{ref_pocus2}. This need for expert radiologists creates a bottleneck in the COVID-19 screening pathway. Typically, POCUS devices are mobile and do not have large computational resources available on-board mandating computationally light-weight solutions. In addition, POCUS analysis is typically performed on video rather than still images. Thus any solution must also operate in real-time.

Motivated by this challenge, this study introduces COVID-Net US, a novel, highly customized self-attention deep neural network architecture tailored specifically for the detection of COVID-19 cases from POCUS images.  The goal with COVID-Net US is to design a highly efficient yet high performing deep neural network architecture that is small enough to be implemented on low-cost devices allowing for limited additional resources needed when used with POCUS devices in low-resource environments. More specifically, we employ a collaborative human-machine design approach. We leverage human knowledge and experience to design an initial prototype network architecture. This prototype is subsequently leveraged alongside a set of operational requirements catered around low-power, real-time embedded scenarios to generate a unique, highly customized deep neural network architecture tailored for COVID-19 case detection from POCUS images via an automatic machine driven design exploration strategy. This combination of human and machine driven design strategy results in a highly accurate yet efficient network architecture. Finally, to demonstrate the efficacy of the proposed COVID-Net US for edge devices, we evaluate its practical performance by deploying it on a Raspberry Pi with a 1.5 GHz ARM Cortex 72 CPU with 4GB memory.  To advocate affordable healthcare and artificial intelligence for resource-constrained environments, we have made COVID-Net US open source and publicly available as part of the COVID-Net open source initiative\footnote{http://www.covid-net.ml/} \cite{ref_covid_net,ref_covid_ct,wong2021covidnet,gunraj2021covidnet} for accelerating
the advancement and adoption of deep learning for tackling this pandemic.
  \vspace{-0.07in}
\section{Related Work}
  \vspace{-0.07in}
There are a multitude of initiatives aiming to apply machine learning to classification of medical images for screening of COVID-19 infections. One such initiative is the COVID-Net initiative, which has shown success in curating open-source publicly available CXR and CT image datasets to act as training data for deep neural network models, and in using machine-driven design exploration to optimize both micro and macroarchitecture to build an architecture tailored to the problem at hand, improving both efficacy and efficiency \cite{ref_covid_net,ref_covid_ct,wong2021covidnet,gunraj2021covidnet}.

There have been numerous other works developing deep neural network architectures for processing POCUS images, especially in the context of COVID-19 infections in recent times. The work of Arntfield et al. showed effectiveness in building deep neural network architectures, using Xception architectures, to identify the presence of COVID-19 in clinically captured lung ultrasound (LUS) images, reaching a  higher area under the receiver operating characteristic curve (AUC) with the developed model than was able to be done by expert radiologists \cite{ref_arntfield}. The authors have released the code used in this research, however neither the data used to train the model nor any resulting models are available, preventing complete reproducibility or even application by others at this time.

In another study \cite{ref_pocovidnet}, the authors constructed a deep neural network architecture trained using their own compiled datasets, which were a mix of public and private datasets, and were able to achieve relatively high performance ($>$0.94 AUC for each of the normal, pneumonia, and COVID-19 classes) with their final model based off of VGG-16. However, the authors did not appear to investigate how the resulting architecture fares on low-cost devices which could limit the contexts in which their work is applicable.

The study by Rojas-Azabache et al.\cite{ref_peru_rpi} involved the development their own deep neural network architecture, trained on POCUS images collected from individuals in the city of Lima, Peru (and kept private by the authors). The authors demonstrate the efficacy of their VGG-16 inspired model not only in identifying COVID-19 but also by deploying it on a Raspberry Pi . Although the study does not report the memory footprint or latency of the proposed model, a standard VGG-16 architecture consists of 134 million parameters and requires 15 billion FLOPS for inference \cite{ref_vgg}. While it is possible to run this model on a Raspberry Pi, the inference will be far from real-time. The protocol for physician analysis of an ultrasound examination consists of them analyzing the video taken by the POCUS device, often from various views and positions \cite{ref_pocus1}. It is likely that for effective automatic analysis, many frames from a single video, or possibly multiple, will need to be processed by a given architecture, so to be feasibly implemented it will need to perform this analysis in as close to real time as possible. Thus, a goal of this study is to develop the architecture such that not only is it effective in terms of the accuracy metrics but also efficient both in size of the resulting architecture and efficiency of model processing time to facilitate affordable use on low-cost embedded processors to better facilitate for widespread use in highly affected regions around the world.
  \vspace{-0.07in}
\section{Methods}
  \vspace{-0.04in}
This study introduces COVID-Net US, a tailored, highly efficient self-attention neural network design that aims to detect COVID-19 patient cases from POCUS images. As mentioned earlier, POCUS systems are typically resource constrained but also mandate real-time solutions. Thus any practical solution would need to exhibit low latency and high accuracy, simultaneously. To this end, this study leverages a machine-driven design exploration to automatically discover highly customized micro and macro architecture designs that result in a high-performing yet highly efficient deep neural network architecture catered around task and operational requirements at hand. The found macro-architecture design take advantage of self-attention via attention-condensers to yield an efficient and accurate network architecture \cite{ref_attendnets}.
  \vspace{-0.07in}
\subsection{COVIDx-US dataset}
  \vspace{-0.07in}
We employ the COVIDx-US dataset \cite{ref_covidxus} to train and evaluate our network. The COVIDx-US dataset contains a total of 173 ultrasound lung videos and 16,822 processed images, curated from multiple sources. The videos comprise of 60 COVID-19 positive cases, 15 normal cases, 41 patients with non-COVID related pneumonia, and 57 "other" non-COVID cases containing other pulmonary issues such as Chronic Obstructive Pulmonary Disorder (COPD), Pneumonthorax, and Pulmonary Oedema. Since the COVIDx-US dataset is compiled from various sources, many of which did not have extensive demographic data available, a complete summary of demographics for the overall dataset is not able to be provided at this time. Of those for which demographic information is known, there are POCUS videos from both male and female individuals with ages ranging from 7 to 72.

\begin{figure*}[!t]
  \centering
  \includegraphics[width= \linewidth]{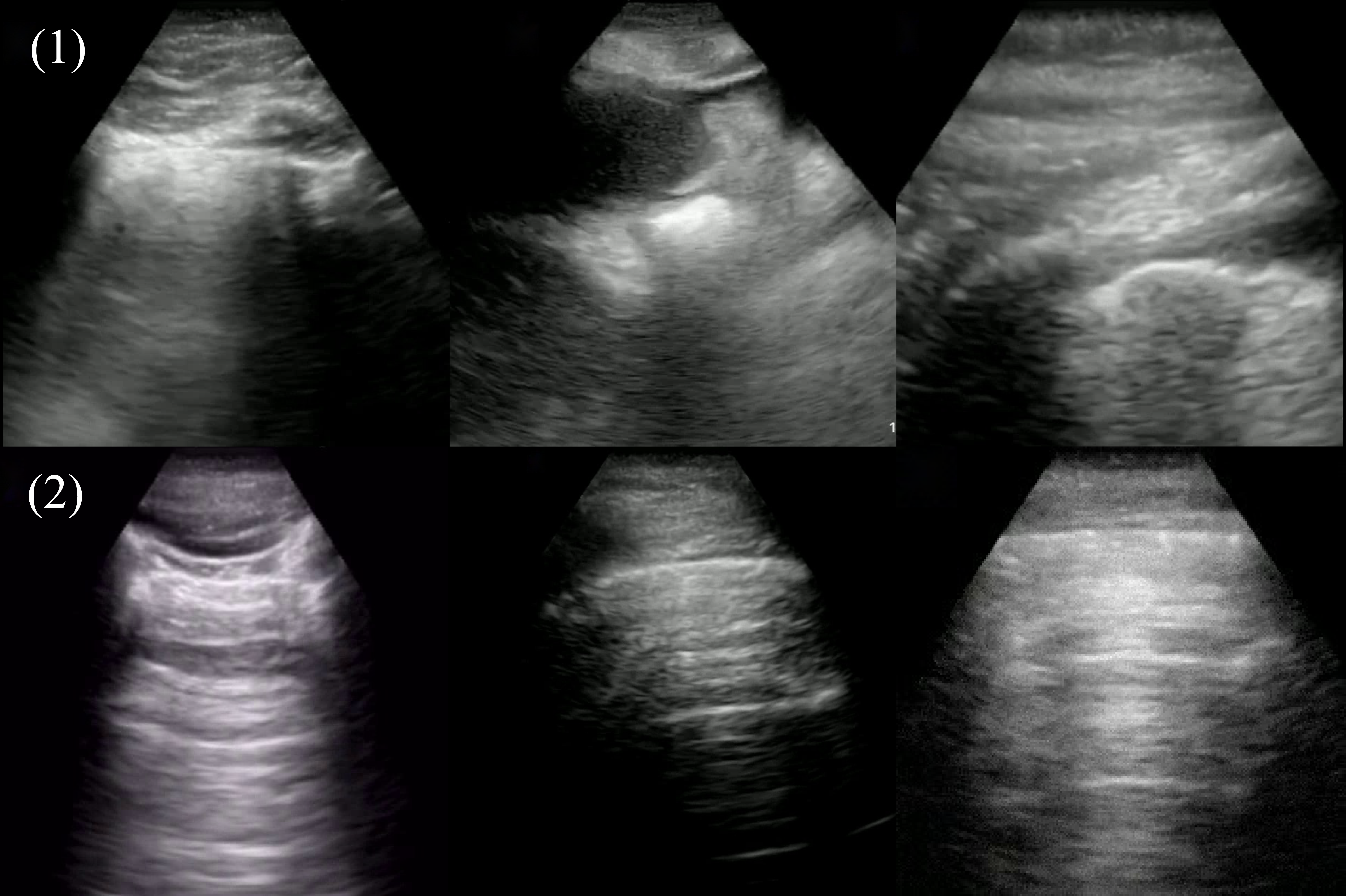}
  \caption{Example POCUS images: (1) SARS-CoV-2 positive patient cases and (2) SARS-CoV-2 negative patient cases.}
  \label{fig:us-examples}
    \vspace{-0.015in}
\end{figure*}

Due to the low number of normal cases and the heterogeneity of the "other" cases, initial attempts for 4-way multi-label classification were challenging. Instead, the dataset was treated as a binary classification problem, where the COVID-19 cases were labeled as positive and the normal, pneumonia, and "other" cases were labeled as negative. Additionally, due to very low numbers of positive examples taken with linear ultrasound probes, only cases of studies with a convex ultrasound probe were selected for this experiment. Selected example images are presented in Figure \ref{fig:us-examples}. The resulting images were split into a training set with 7,644 images (3,947 positive and 3,697 negative), a validation set with 2,383 images (1,085 positive and 1,298 negative), and a test set with 2,233 images (1,052 positive and 1,181 negative). The splits were made such that all frames from each video were either in the training, validation, or testing set, and patient-level demographic information was leveraged in a similar manner, ensuring that obvious instances of data leakage across sets is avoided to the greatest possible extent.

  \vspace{-0.07in}
\subsection{Network Design}
  \vspace{-0.07in}
In this study, we construct an initial network design prototype by leveraging residual architecture design principles~\cite{ref_resnet} given their ability to achieve high performance while maintaining ease of training by alleviating the issue of vanishing gradient present in conventional deep neural network architectures. More specifically, the initial network design prototype was designed to provide one of two predictions: a) SARS-CoV-2 negative c) SARS-CoV-2 positive.

Based on this initial network design prototype and a set of operational requirements catered around an embedded scenario, we leverage a machine-driven design exploration strategy to automatically determine the micro and macro-architecture for the final COVID-Net US deep neural network. This strategy allows us to automatically strike a balance between detection accuracy and network efficiency, yielding a network design that is tailored for on-device detection of COVID-19 from POCUS images.

In recent years, there has been increased interest in automatic exploration of the architectural space to yield tailor-made network designs that achieve both excellent accuracy while simultaneously having a small memory footprint and and an efficient architecture. Methods such as MONAS \cite{ref_monas} and DARTS \cite{ref_darts} employ reinforcement learning or gradient descent to find an optimal set of parameters and thus yield an architecture that satisfies the given operational requirements and constraints. In this study, we leverage generative synthesis \cite{ref_gensys} to explore micro and macro-architecture designs in an efficient yet effective manner. In contrast to techniques that rely on gradient descent or reinforcement learning, generative synthesis finds the optimal network architecture through an iterative process that solves a constrained optimization problem. More formally, generative synthesis can be formulated as the following expression

\begin{equation}
\mathcal{G}=\max_{\mathcal{G}}\mathcal{U}(\mathcal{G}(s)) \;\;\;\textrm{ $subject to \;\;\; 1_r(G(s))=1, \;\;\forall \in \mathcal{S}$}.
\end{equation}

where the goal is to find generator $\mathcal{G}$ that generates deep neural network architectures that maximize a performance function $\mathcal{U}$ (e.g. \cite{ref_netscore}) given the operational constraints $1_r(\cdot)$. Generative synthesis aims to find an approximate solution to the above expression through an iterative process. The ultimate goal of this study is to produce an efficient and accurate network architecture that can be employed on ultrasound machines and detect COVID-19 in real-time. To this end, we formulate $1_r(\cdot)$ in terms of parameter count ($<$ 1M), number of FLOPS ($<$ 1B) and area under the curve (AUC) ($>$0.9).

\begin{figure*}[!t]
  \centering
  \includegraphics[width= \linewidth]{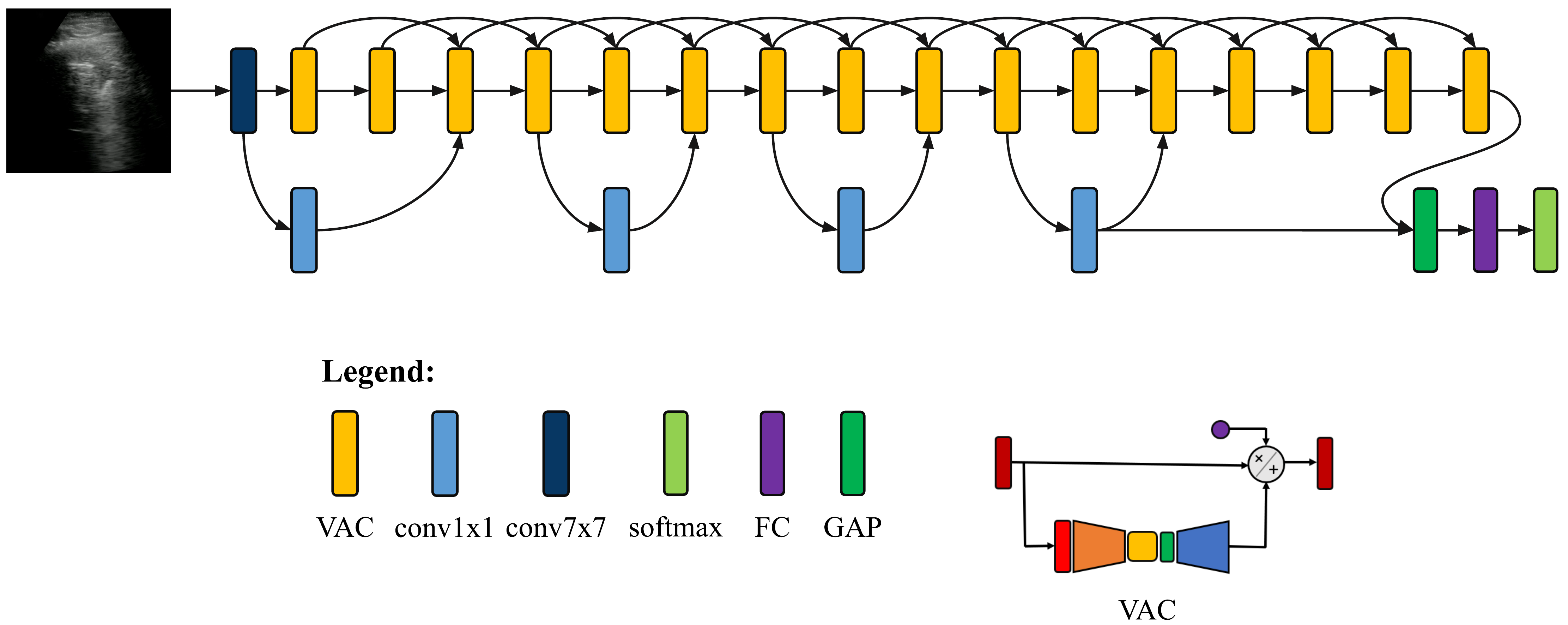}
  \caption{The COVID-Net US architecture design. The COVID-Net US design exhibits high architectural heterogeneity and the use of visual attention condensers with macro-architecture and micro-architecture designs tailored specifically for the detection of COVID-19 from ultrasound images.}
  \label{fig:arch}
  \vspace{-0.2in}
\end{figure*}

Figure~\ref{fig:arch} describes the resulting network. Several interesting observations can be made from the resulting architecture. First, we can note that that network design is comprised of a different number of macro-architectures including pointwise convolutions, attention condensers and standard convolutions. This high macro-architecture diversity is a direct result of using machine-driven design strategy to identify an architecture design specifically for identification of COVID-19 in ultrasound images. Secondly, we note that the use of depthwise convolutions leads to a very computationally light-weight design. This light-weight design is the result of imposing constraints on the parameter count and the number of FLOPS. Finally, we note that the architecture relies heavily on attention condensers. Attention condensers were originally introduced in \cite{ref_attendnets} and have demonstrated themselves to be a highly efficient mechanism for self-attention \cite{ref_selfattn}. Visual attention condensers produce a compressed embedding that represents the spatial and cross-channel activation relationships. This compressed representation yields selective attention which improves representational capability while exhibiting low architectural complexity.

  \vspace{-0.07in}
\subsection{Explanation-driven Performance Validation}
  \vspace{-0.07in}
To further validate the performance of the COVID-Net US, we examine the decision-making behaviour of the resulting architecture by employing GSInquire, a state-of-the-art explainability method which identifies critical factors in the test data which are quantifiably shown to be integral to the decisions made by the system \cite{ref_gsinquire}.  In short, GSinquire makes use of an inquisitor $\mathcal{I}$ in a generator-inquisitor pair ${\mathcal{G}, \mathcal{I}}$ during the machine-driven exploration generative synthesis described above \cite{ref_gensys}. During generative synthesis, the inquisitor $\mathcal{I}$ probes the connections in the network $\mathcal{N}$ (here, COVID-Net US) iteratively with the generator $\mathcal{G}$ to identify the factors in the input to the network that are affecting the decision-making process of $\mathcal{N}$. The result of this iterative process is a highlighting of the regions or components of desired test inputs, in this case, test images, which if removed or modified would most drastically change the decision made by the network $\mathcal{N}$ and thus are critical factors.

In the context of COVID-Net US, the results of this explanation-driven performance validation is to identify whether the architecture's decision-making process is successfully using the factors in ultrasound examinations noted by radiologists as important in identifying the presence or absence of COVID-19 in individuals. To that end, the critical factors will be compared to those noted in the previously mentioned studies \cite{ref_pocus1} and \cite{ref_pocus2} to validate that the architecture makes use of clinically relevant features interpretable by radiologists.

Select patient cases from the benchmark dataset are also reviewed and reported on by a practicing clinician (20 years of clinical practice) specializing in intensive care (ICU) and 15 years of expertise in POCUS interpretation.
  \vspace{-0.07in}
\section{Results and Discussion}
  \vspace{-0.07in}
We evaluate the resulting network architecture through a number of important metrics. Historically, machine learning has been mostly concerned with providing accurate performance, which is still vital in our case, and to that end the architecture with the AUC score to keep consistent with previously mentioned studies \cite{ref_arntfield,ref_pocovidnet}. Additionally, to ensure that the resulting architecture meets the requirements of effectiveness on low-resource, "edge" devices, the feasibility of this implementation will be evaluated by comparing the size and speed of processing on a test device, a Raspberry Pi.  For comparison purposes, we also evaluate the performance of a ResNet-50~\cite{ref_resnet} architecture design. We also perform qualitative analysis using an explainability-driven performance validation method.
  \vspace{-0.07in}
\subsection{Quantitative Analysis}
  \vspace{-0.07in}
To further contextualize the balance of an architecture's performance compared to its computational and architectural complexities, which predict well its feasibility of implementation in realistic contexts, researchers have proposed many quantitative metrics, of which NetScore is notable \cite{ref_netscore}. NetScore is based on an analysis of many publicly available deep neural network architectures, and is described as useful in evaluating the effectiveness of architectures in practical contexts on edge devices, which describes well the use case of analyzing POCUS images in low-resource situations.

\begin{table}[]
\centering
\caption{Comparison of test AUC, FLOPS and model size of the proposed COVID-Net US architecture with the ResNet-50 architecture.}
\label{tab:results_table}
\begin{tabular}{lccccc}
\textbf{Model} & {\textbf{~~NetScore~~}} & {~~\textbf{Params}~~} & {~~\textbf{FLOPS}~~} & {~~\textbf{Latency (ms)}~~} & {\textbf{Test AUC}}  \\
\midrule
COVID-Net US  &   82.53   &   65K    & 596M    & 216      & 0.9824  \\
ResNet-50    &   49.08   &   23M   & 37B     & 3087    & 0.9196   \\ \bottomrule
\end{tabular}
\end{table}

Table \ref{tab:results_table} presents the accuracy results and the architecture and computation complexity results (NetScore, number of parameters, FLOPs, and Latency in ms on a Raspberry Pi).  In terms of performance on the desired task, namely the AUC of the resulting model during testing, both the ResNet-50 and the COVID-Net US network architecture designs performed well with both achieving AUC values higher than 0.9, with COVID-Net US reaching above 0.98, comparable to the related works. Since the datasets used by each work are different, comparison of the exact performance values across studies is difficult, and in future work, studies that release their training code such as \cite{ref_arntfield} and \cite{ref_pocovidnet} will be tested to better compare the results of our training methods and analyze the sensitivity and positive predictive value of the architectures. As such, the effectiveness of the resulting architecture is consistent with the related works in suggesting that automated detection systems can be successful in identifying cases of COVID-19 infections from ultrasound examination, and thus prove effective as a complementary COVID-19 screening tool, particularly where RT-PCR tests or CXR or CT examinations could be difficult due to resource or time constraints.

The second level of analysis concerns the measures of computational and architectural complexity. As desired, the utilization of a machine-driven design exploration strategy to determine the best micro and macro-architecture designs resulted in the proposed COVID-Net US outperforming ResNet-50 across all metrics in this area. The NetScore achieved by COVID-Net US was 82.53 compared to 49.03 as achieved by ResNet-50; in the original study, examined models ranged from around 30 to 80 on this scale, showing that this difference is significant. COVID-Net US exhibits various macro-architectures which work in tandem for greatly reduced architectural and computational complexity. The result is a highly efficient architecture that requires just 65K parameters (\textbf{353$\times$} fewer than ResNet-50), 596M FLOPS (\textbf{62$\times$} lower than ResNet-50) and exhibits an on-device latency of just 216 ms (\textbf{14.3$\times$} lower than ResNet-50) on a Raspberry Pi. The small size and low latency of the proposed architecture enable them to be deployed on POCUS devices without significant hardware cost.
  \vspace{-0.07in}
\subsection{Qualitative Analysis}
  \vspace{-0.07in}
\begin{figure*}[!t]
  \centering
  \includegraphics[width= \linewidth]{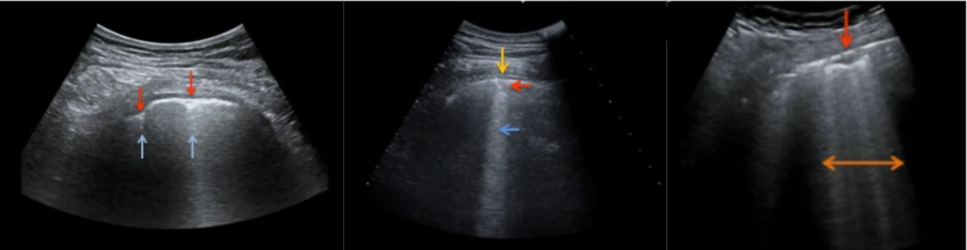}
  \caption{Examples of ultrasound images with annotations of signs used by radiologists for scoring the examinations, on a scale from 0 to 3 with 0 being normal and 3 being the most severe \cite{ref_pocus2}. A score 1 example is shown on the left, with red arrows pointing to indentations in the pleural line. The central image shows a score 2 example with breaking points in the pleural line annotated, as well as "white lung" artefacts extending further down. Score 3 is shown in the right image, with significantly discontinuous pleural lines with a large area of white lung.}
  \label{fig:radiologist_pocus_signs}
    \vspace{-0.15in}
\end{figure*}

\begin{figure*}[!t]
  \centering
  \includegraphics[width= \linewidth]{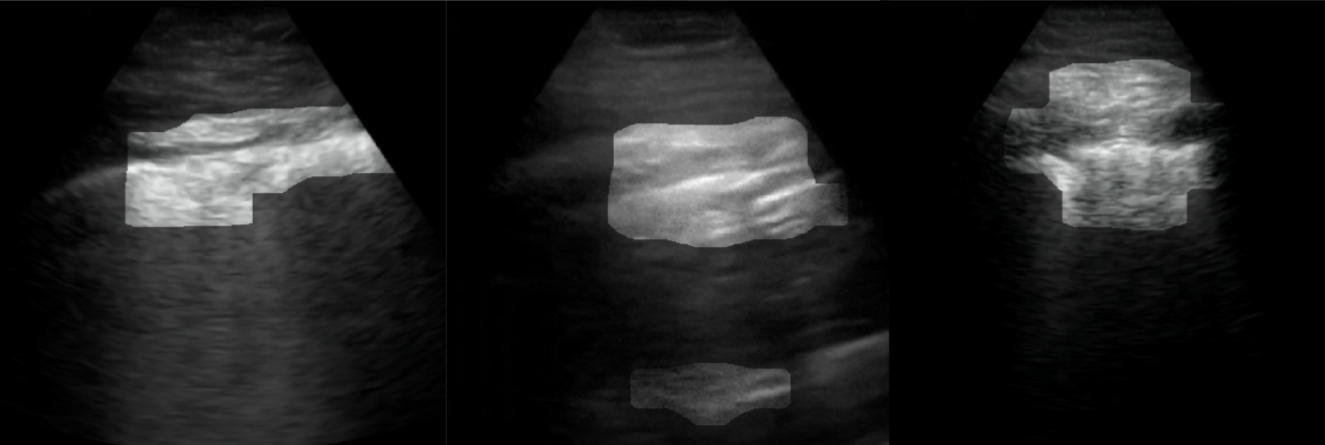}
  \caption{Examples of COVIDx-US test images with GSInquire critical factor results overlaid, each for separated positive COVID-19 cases. Each image has the pleural line region identified as important to the decision making process, and many have regions further down the lung as highlighted as well, where white lung would be found.}
  \label{fig:gsinquire_results}
    \vspace{-0.15in}
\end{figure*}

As described previously, explainability-driven performance validation was performed on COVID-Net US using GSInquire. Figure \ref{fig:radiologist_pocus_signs} shows examples of images corresponding to scores 1, 2, and 3 (out of 3, with 3 being more severe cases of lung involvement) based on the system described by \cite{ref_pocus2}. Figure \ref{fig:gsinquire_results} then shows examples of the output from GSInquire for selected test images from COVIDx-US, with the critical factor regions highlighted. Note that the regions identified as quantifiably important for COVID-Net US when making classifications line up well with the areas annotated by radiologists based on their experiences. While this does not prove that COVID-Net US is making the exact same decisions as those radiologists, is does suggest two things; first, it confirms that the architecture is learning to examine features within the region of interest of the images, rather than unrelated regions or possible artefacts beyond the extent of the relevant content; and second, that since the architecture is focusing on the same regions as radiologists it is likely making clinically relevant decisions helping clinicians to trust the results of those decisions, supporting the notion that such models can be relied upon to automate the screening process for COVID-19 using ultrasound images.

The clinical expert findings and observations with regards to the critical factors identified by
GSInquire for select patient cases shown in Figure \ref{fig:gsinquire_results} are as follows. In all these cases, COVID-Net US detected them to be SARS-CoV-2 positive, which was clinically confirmed. It was observed in all three cases that the critical factors leveraged by COVID-Net US correspond to pleural line irregularities, specifically hypoechoic areas and breakages in the normally smooth contoured pleural surface (subpleural consolidations) associated with lung pathology consistent with SARS-CoV-2 induced lung inflammation and pneumonia. As such, it was demonstrated that the identified critical factors leveraged by COVID-Net US are consistent with radiologist interpretation.
  \vspace{-0.07in}
\section{Conclusions}
  \vspace{-0.07in}
In this study, we introduce COVID-Net US, a highly efficient, self-attention based deep convolutional neural network architecture tailored for screening of COVID-19 infections from lung ultrasound images in low-resource environments. A machine-driven exploration strategy was used to perform the architecture design, resulting in an architecture that made use of visual attention condensers highly customized to the task at hand. Experimental results show that COVID-Net US achieved a higher accuracy than the ResNet-50 in terms of AUC score, while having significantly lower architectural and computational complexity, making it much more feasible of an architecture in low-resource contexts. Additionally, explainability-driven performance validation was done to ensure that the decision-making process of COVID-Net US is consistent with that of COVID-19 lung ultrasound examination protocols published by radiologists.

With the expectation that COVID-19 will continue to have an impact on global health for the foreseeable future, it is important that work done to combat the pandemic is ongoing in all possible avenues. The ability for COVID-Net US to allow for point-of-care ultrasound devices to act as a COVID-19 screening tool with less of a need for radiologist oversight for analysis means that contexts where current commonly used tools, namely RT-PCR, CXR, and CT, are rare or unavailable due to resource constraints can hopefully improve their ability to screen for the disease. Whether that is in contexts with time constraints, such as emergency situations, or with resource constraints, such as developing countries, they will be able to benefit from this tool and protect individuals from further detrimental effects. We look forward to this work expanding the capabilities of the general COVID-Net initiative to provide support to clinicians and researchers alike.
  \vspace{-0.07in}
\section*{Acknowledgements}
  \vspace{-0.07in}
We thank the Natural Sciences and Engineering Research Council of Canada (NSERC), the Canada Research Chairs program, DarwinAI Corp., and the National Research Council Canada (NRC). The study has received ethics clearance from the University of Waterloo (42235).

\bibliographystyle{splncs04}
\bibliography{references}

\end{document}